\documentclass[DIV=calc, paper=a4, fontsize=8pt, twocolumn]{scrartcl}

\usepackage{graphicx}
\usepackage{comment}
\usepackage[cmex10]{amsmath}
\newcommand{\A}{\mathcal{A}}
\newcommand{\D}{\mathcal{D}}
\newcommand{\G}{\mathcal{G}}
\newcommand{\N}{\mathcal{N}}
\newcommand{\M}{\mathcal{M}}
\newcommand{\B}{\mathcal{B}}
\newcommand{\C}{\mathcal{C}}
\begin{document}
\title{Energy Coupled Mode Theory for Electromagnetic Resonators}
\author{Sameh~Y.~Elnaggar,~Richard~J.~Tervo, Saba~M.~Mattar}
\date{}       
\maketitle

\begin{abstract}
There is recent interest in the inter/intra-element interactions of metamaterial unit cells. To calculate the effects of these interactions which can be substantial, an \emph{"ab-initio"} general coupled mode equation, in the form of an eigenvalue problem, is derived. The solution of the master equation gives the coupled frequencies and fields in terms of the uncoupled modes. By doing so, the problem size is limited to the number of modes rather than the, usually large, discretized spatial and temporal domains obtained by full-wave solvers. Therefore, the method can be considered as a \emph{numerical recipe} which determines the behavior of a complex system once its simpler \emph{ingredients} are known. Besides quantitative analysis, the coupled mode equation proposes a pictorial view of the split rings' hybridization. It can be regarded as the electromagnetic analog of molecular orbital theory. The solution of the eigenvalue problem for different configurations gives valued information and insight about the coupling of metamaterials unit cells. For instance, it is shown that the behavior of split rings as a function of the relative position and orientation can be systematically explained. This is done by singling out the effect of each relevant parameter such as the coupling coefficient and coupled induced frequency shift coefficients.

\end{abstract}

\section{Introduction}

METAMATERIALS (MMs) are artificial media that exhibit fascinating electromagnetic properties not present in nature \cite{Pendry99, Pendry00, Veselago68}. They are usually manufactured from arrays of sub-wavelength resonators. Therefore, effective media theory is used to study their behavior \cite{Pendry99, Smith05}. By analogy with natural materials, the sub-wavelength resonators form the basic unit cells (atoms) of the MM. Each unit cell has electric and magnetic multipoles. If the cells inter-spacing is small, they can strongly interact and hence substantially change the media properties \cite{Liu09, Powell10, Liu10}. Inspired by stereochemistry, the interaction can be controlled by changing the spatial arrangement of the resonators. This concept, widely known as \emph{stereometamaterials}, was applied to different MM configurations \cite{Liu09,Liu10, Ekmekci11, Keiser13, Singh10, Lapine12}. Similarly, the media properties can change by tuning the unit cell's resonant frequency. This can be done for example, by incorporating a photo-conductive semiconductor \cite{Kafesaki12, Keiser13}.

The analogy between the MM unit cells and atoms is utilized to visualize the interaction between MM unit cells as the \emph{hybridization} of unit cells' modes. This concept was first applied to the study of nano-shells and nano-spheres \cite{Prodan03, Prodan04}. Later, it was adopted to qualitatively and experimentally analyze the interaction between either split ring resonators (SRRs), which are the building blocks of MM \cite{Guo07, Aydin10, Lahiri11, Yang12}. The interaction between the dipole moments determines the nature and strength of coupling. Due to the bi(iso/aniso)tropic property of split rings, both electric and magnetic dipoles play roles in the process of coupling \cite{Marques02}.

Because of the sub-wavelength nature of MM unit cells, the coupling between them was studied based on quasi-static approaches. For example, the current and charge densities are used to determine the Lagrangian ($\mathcal{L}$). The Lagrangian equation of motion yields a system of coupled differential equations which determines the interaction terms \cite{Powell10, Liu09, Sersic09}. Because of their importance, the analysis is usually carried out for meta-dimers. It is shown that the net effect of both the in-plane electrical and the out-of-plane magnetic dipoles determines the interaction strength \cite{Sersic09, Liu10}. Circuit models are also developed to model and to quantify split rings and their coupling \cite{Baena05, Shamonina05, Poo14}.

Coupled Mode Theory (CMT) proves to be a very successful tool when applied to weakly coupled systems \cite{Haus90}. It was used in MM \cite{Nguyen07, Sukhourkov14}, microwave filters \cite{Amari10, Awai06, Awai08, Levie13}, wireless power transfer \cite{Karalis08, Kurs07, Sample11, Xiaofang11} and magnetic resonance \cite{Mattar11, Paper4, Paper5, Paper6}. Unlike finite element, finite difference and method of moments, CMT reduces the computational domain to the number of modes. Other than reducing the computational complexity, CMT provides an intuitive picture of how a complex system behaves in terms of the interaction of its relatively simpler subsystems. This makes CMT a very useful and powerful tool for studying MM.

With the aim of qualitatively and quantitatively analyzing the hybridization of coupled systems including MMs, an \emph{"ab-initio"} coupled mode formalism is developed in the current article. A similar analysis was developed to study an electron paramagnetic probe consisting of a dielectric resonator and a cavity. However, it was limited to the studied case and bounded to TE modes \cite{Paper4}. Nevertheless, it was shown that a coupled mode formalism is still capable of describing the system behavior even though the coupling coefficient can be substantially large ($\approx 0.4$). Besides the ability to calculate the eigen-frequencies, field dependent parameters such as the quality factor and resonator efficiency were accurately determined \cite{Paper5, Paper6}. The current article provides a systematic derivation of the general coupled mode equation in the form of an eigenvalue problem. Once solved, the eigenvalues determine the resonant frequencies (eigen-frequencies), while the eigenvectors are used to find the fields. The eigenvalue equation is proven to obey the energy conservation principle and hence is named energy coupled mode theory (ECMT). It reduces to well known formulae when applied to special cases, such as the hybridization of meta-dimers \cite{Powell10}. Moreover, ECMT provides a complimentary approach to the dipoles coupling widely used in examining the interaction of SRRs. It also gives a numerical procedure for calculating frequencies and fields. Two double SRRs (DSRRs) configurations are numerically studied. The results are explained by assessing the effect of system parameters on the coupling coefficient $\kappa$.

The paper is partitioned as follows: Section II presents the theoretical background with emphasis on field expansions and notations used. Section III is devoted to the theoretical derivation. Section IV presents the results and discusses the hybridization of split-ring resonators. Finally, the conclusion follows in Section V.

\section{Theoretical Background}

The fields of a system of coupled resonators are expanded in terms of the fields of the uncoupled subsystems, which are regarded as a basis set and they are not necessarily orthogonal. In general this set is infinite. For practical purposes, it can be truncated to a finite one of a suitable size $N$. Therefore

\begin{equation}
\label{EfieldExp}
\textbf{E}=\sum\limits_{i=1}^N a_{i} \textbf{E}_i
\end{equation}
and
\begin{equation}
\label{HfieldExp}
\textbf{H}=\sum\limits_{i=1}^N b_{i} \textbf{H}_i.
\end{equation}

Here $a_i$ and $b_i$ are the expansion coefficients of the electric and magnetic components respectively. The $N$ modes can be equal to or greater than the number of resonators. So a valid coupled mode can be the linear combination of the first mode of resonator one and the first mode of resonator two, or it can be the first mode of resonator one, the second mode of resonator one and the first mode of resonator two, etc.  This procedure is very helpful, for instance, if the frequencies of two modes of one resonator are very close (or degenerate), so they both couple with a third nearby mode of a different resonator. Expansions (\ref{EfieldExp}) and (\ref{HfieldExp}) are equivalent to the linear combination of atomic orbitals (LCAO) in molecular orbital theory \cite{Silbey05}.

At the conductors' surfaces, the surface current density $\textbf{J}_s$ is equal to $\hat{n}\times\textbf{H}$ where $\hat{n}$ is the unit normal. Therefore, the current density $\textbf{J}$ is expanded in terms of the uncoupled fields as
\begin{equation}
\label{JExpand}
\textbf{J}=\sum\limits_{i=1}^N b_{i} \textbf{J}_i.
\end{equation}

Each $i^{th}$ uncoupled mode satisfies the sinusoidal time-varying Maxwell’s equations \cite{Pozar05}, therefore the curl of the fields can be written as:

\begin{equation}
\nabla\times \textbf{E}_i=-j\omega_i\mu_i(\textbf{r}) \textbf{H}_i,
\end{equation}

\[
\nabla\times \textbf{H}_i=j\omega_i\epsilon_i(\textbf{r}) \textbf{E}_i+\textbf{J}_i.
\]

Here $\omega_i, \mu_i , \epsilon_i$ and $\textbf{J}_i$ are the angular frequency ($2\pi f_i$), permeability, permittivity and current density of the $i^{th}$ uncoupled resonator respectively. In general, $\epsilon_i$ and $\mu_i$  change with position $\textbf{r}$.  Similarly, for the coupled system
\[
\nabla\times \textbf{E}=-j\omega\mu(\textbf{r}) \textbf{H}
\]

\begin{equation}
\nabla\times \textbf{H}=j\omega\epsilon(\textbf{r}) \textbf{E}+\textbf{J}
\end{equation}
where $\omega, \mu , \epsilon$  and $\textbf{J}$ are the corresponding symbols for the coupled system.

In the current article, the Dirac Bra-ket notation is used to represent the inner product. For example, the inner product of two vector fields$\textbf{A}$ and $\textbf{B}$ is denoted by $\langle \textbf{A}|\textbf{B}\rangle$ and by definition it is equal to
\[
\langle \textbf{A}|\textbf{B}\rangle \equiv \int_V \textbf{A}^*\cdot\textbf{B}\ dv,
\]
where $V$ is the total volume.
\section{Analytical Derivation}

\subsection{Resonance Condition}
In this subsection, the resonance condition of the coupled system is derived in terms of the uncoupled parameters. For practical resonators, the losses are small and ignored. Later on after determining the fields, the losses can be obtained \cite{Pozar05, Paper5}. Therefore, the time-averaged Poynting vector $\tilde{\textbf{R}} = \textbf{E} \times \textbf{H}^* $ and $\textbf{E}\cdot\textbf{J}^*$ are imaginary and the complex power equation is written as
\begin{equation}
\label{Poynting}
{\frac{1}{2}\oint_{\partial V}\textbf{E}\times\textbf{H}^*\cdot d\textbf{S}}+2\omega\int_V (\tilde{W}_M-\tilde {W}_E)dv=-\frac{1}{2} {\langle \textbf{J} | \textbf{E} \rangle}
\end{equation}
Here $\tilde {W}_M$ and $\tilde {W}_E$ are the time-average stored magnetic and electrical energy
\begin{equation}
\label{WM}
\tilde{W}_M=\frac{1}{4}\langle \mu \textbf{H} | \textbf{H} \rangle
\end{equation}
\begin{equation}
\label{WE}
\tilde{W}_E=\frac{1}{4}\langle \epsilon \textbf{E} | \textbf{E} \rangle
\end{equation}
respectively. At resonance
\begin{equation}
\label{Resonance}
\tilde {W}_M=\tilde {W}_E.
\end{equation}
The Power equation (\ref{Poynting}) is simplified to be
\begin{equation}
\label{PoySimp}
\oint_{\partial V}\textbf{E}\times\textbf{H}^*\cdot d\textbf{S}+\langle \textbf{J}|\textbf{E}\rangle=0.
\end{equation}

By using expansions (\ref{EfieldExp}), (\ref{HfieldExp}), (\ref{JExpand}) and taking the complex conjugate, (\ref{PoySimp}) is written as

\begin{equation}
\label{MatResonance}
a^\dag (\M+\N)b=0.
\end{equation}
Here $\M_{ik}=\int_{\partial V} \textbf{E}_i^*\times\textbf{H}_k\cdot d\textbf{S}$ and $\N_{ik}=\langle\textbf{E}_i|\textbf{J}_k\rangle$.
The integral factors $\M_{ik}$ and $\N_{ik}$ in (\ref{MatResonance}) can be expressed in terms of the stored energy quantities by expanding $\nabla\cdot(\textbf{E}_i\times\textbf{H}_k^*)$, using the identity

\begin{equation}
\nabla \cdot (\textbf{A}\times\textbf{B})=(\nabla\times\textbf{A})\cdot\textbf{B}-(\nabla\times\textbf{B})\cdot\textbf{A}
\end{equation}
and integrating over the total volume $V$
\begin{multline}
\label{useful}
\oint_{\partial V}\textbf{E}_i\times\textbf{H}_k^*\cdot d\textbf{S}=-j\omega_i\langle\mu_i\textbf{H}_i|\textbf{H}_k\rangle+j\omega_k\langle\epsilon_k\textbf{E}_k|\textbf{E}_i\rangle-\langle \textbf{E}_i|\textbf{J}_k\rangle
\end{multline}

Relation (\ref{useful}) can be re-written as
\begin{equation}
\label{useful2}
\M_{ik}+\N_{ik}=j\omega_i\B_{ik}-j\omega_k\D^*_{ki},
\end{equation}
where $\B_{ik}\equiv\langle\mu_i\textbf{H}_i|\textbf{H}_k\rangle$ and $\D_{ik}\equiv\langle\epsilon_i\textbf{E}_i|\textbf{E}_k\rangle$.
Because $\B_{ik}$ and $\D_{ik}$ depend on the uncoupled parameters, they are called the uncoupled magnetic and electric energy respectively. The full matrix expression of (\ref{useful2}) is
\begin{equation}
\label{UsefulMat}
\M+\N=j\Omega \B-j\D^\dag\Omega ,
\end{equation}
where
\begin{equation}
\Omega=\left ( \begin{array}{cccc}
 \omega_1 & 0 &\cdots\cdots&0 \\
 0 & \omega_2&...&0 \\
 \vdots &\vdots&\ddots &\vdots\\
 0 & 0 &\cdots&\omega_n
\end{array}\right )
\end{equation}

Equations (\ref{useful2}) and (\ref{UsefulMat}) relate the complex conjugate of the reactive power components $M_{ik}$ and  $N_{ik}$ to the bulk stored energy components $\B_{ik}$ and $\D_{ik}$.

\subsection{Eigenvalue Problem}
   To find the eigenvalue equation, the total fields $\textbf{E}$ and $\textbf{H}$ are projected on the $i^{th}$ field components. Projecting the total magnetic field $\textbf{H}$ onto the $i^{th}$ electric field component, $\textbf{E}_i$ one can write
\[
\nabla\cdot(\textbf{E}_i^*\times\textbf{H})=(\nabla\times\textbf{E}_i^*)\cdot\textbf{H}-(\nabla\times\textbf{H})\cdot\textbf{E}_i^*.
\]

Expanding $\textbf{E}$ and $\textbf{H}$ in terms of the fields $\textbf{E}_i$ and $\textbf{H}_i$ of the uncoupled systems, using (\ref{EfieldExp}) and (\ref{HfieldExp}), and integrating over the whole volume,

\begin{equation}
\label{firstCoupled}
\sum\limits_k(\omega_k\D^*_{ki}b_k-\omega \A_{ik}a_k)=0
\end{equation}
where $\A_{ik}=\langle\epsilon\textbf{E}_i|\textbf{E}_k\rangle$ is the $\left(i,j\right)$ coupled electrical energy.

Equation (\ref{firstCoupled}) relates the $a_k$ and $b_k$ coefficients to one another. One needs to find another relation between the two coefficients and substitute $b_k$ in term of $a_k$ to obtain the eigenvalue equation. This can be achieved by projecting the total electric field $\textbf{E}$ onto the $i^{th}$ magnetic field component, $\textbf{H}_i$ and noting that the electric field is either zero at infinity for open structures or it is normal to the bounding surface if the system is enclosed in a shield or a cavity, $\oint_{\partial V}\textbf{E}\times\textbf{H}_k^*\cdot d\textbf{S}=0$. Thus one can arrive at,
\begin{equation}
\label{secondCoupled}
\sum\limits_k -\omega \G_{ik}b_k+\omega_i \D_{ik}a_k=0,
\end{equation}
where $\G_{ik}\equiv\langle \mu\textbf{H}_i|\textbf{H}_k\rangle$ is the $\left(i,j\right)$ coupled magnetic energy.

The two relations (\ref{firstCoupled}) and (\ref{secondCoupled}) can be written in matrix form as
\begin{equation}
\label{firstMatrix}
{\D}^\dag\Omega b-\omega \A a=0,
\end{equation}
\begin{equation}
\label{secondMatrix}
-\omega \G b+\Omega \D a=0.
\end{equation}

These two equations represent the projection of the coupled total fields onto the uncoupled ones. From here, the eigenvalue equation can be derived by noting that from (\ref{secondMatrix})
\begin{equation}
\label{secMatrix2}
b=\frac{1}{\omega}\B^{-1}\Omega \D a.
\end{equation}
Substituting (\ref{secMatrix2}) back in (\ref{firstMatrix}) one arrives at,
\begin{equation}
\label{eigen1}
\left (\A^{-1} \D^\dag\Omega \right)\left(\G^{-1}\Omega \D\right)a=\omega^2 a.
\end{equation}

Equation (\ref{eigen1}) is the required eigenvalue equation, where the eigenvalues are the square of the angular frequency and the eigenvectors are the coefficients of the fields given by (\ref{EfieldExp}). It represents a \emph{numerical recipe} which mixes the \emph{ingredients} (uncoupled modes) in specific \emph{quantities} (determined by the strength of overlap between the fields) to obtain the coupled frequencies and fields.

\section{Results and Discussion}
\subsection{Energy Conservation}
It is interesting to verify that, at resonance, the eigenvalue problem (\ref{eigen1}) satisfies the energy conservation principle given by (\ref{Resonance}) or (\ref{MatResonance}). This is done by expanding the energy expressions (\ref{WM}) and (\ref{WE}) in $\textbf{E}$ and $\textbf{H}$ according to (\ref{EfieldExp}) and (\ref{HfieldExp}). After some algebraic manipulation one can find that,
\[
\tilde{W}_E =\frac{1}{4}a^\dag \A a, \tilde W_M=\frac{1}{4}b^\dag \G b.
\]

Using (\ref{secMatrix2})
\begin{multline}
\tilde{W}_M=\frac{1}{4}b^\dag \G b=\frac{1}{4}(\frac{1}{\omega}\B^{-1}\Omega \D a)^\dag \G(\frac{1}{\omega}\G^{-1}\Omega \D a)\\
=\frac{1}{4}a^\dag(\frac{1}{\omega^2} \D^\dag \Omega \G^{-1}\Omega \D)a.
\end{multline}
The eigenvalue problem (\ref{eigen1})can be rewritten as
\[
\A a=\frac{1}{\omega^2}\D^\dag\Omega \G^{-1}\Omega \D a
\]

Therefore,
\begin{equation}
\label{EnergyProof}
\tilde{W}_M =\frac{1}{4}a^\dag \A a=\tilde{W}_E.
\end{equation}

Relation (\ref{EnergyProof}) is identical to the resonance condition in (\ref{Resonance}). This verifies that the eigenvalues and eigenvectors found using (\ref{eigen1}) guarantee that the system obeys the law of conservation of energy.

\subsection{Hybridization of Double split ring resonators}
In this subsection, the eigenvalue problem (\ref{eigen1}) is solved to find the modes of a DSRR based on the hybridization of the two rings' fundamental modes.  It is worth mentioning that the analysis is applicable to a broad class of interacting resonators such as loop-gap resonators \cite{Hyde89}, degenerate meta-dimers \cite{Powell10}, asymmetric meta-dimers \cite{Keiser13} and DSRRs. Because the rings are thin, all the permittivity functions  $(\epsilon_1(\textbf{r}), \ \epsilon_2(\textbf{r}), \ \epsilon(\textbf{r}),)$ can be approximated by $\ \epsilon_0$. Thus, $\A=\D$ and (\ref{eigen1}) reduces to
\begin{equation}
\label{eigenDSRR}
 \Omega \G^{-1}\Omega \A a=\omega^2a
\end{equation}

If the integrations' volume is large and contains the near fields, then $\G_{ii}=\A_{ii}$. Moreover, the coupling is weak such that $\A^2_{ii}\gg\A_{12}\A_{21} , i=1,2$. Therefore, the eigenvalue problem (\ref{eigenDSRR}) is simplified to
\begin{equation}
\label{Simp1}
\left(\begin{array}{cc}
\omega_1^2 &\omega_1\frac{\omega_1\A_{12}-\omega_2\G_{12}}{\A_{11}}\\
\omega_2\frac{\omega_2\A_{21}-\omega_1\G_{21}}{\A_{22}} & \omega_2^2
\end{array}\right)
\left( \begin{array}{c} a_1 \\ a_2 \end{array} \right)=\omega^2\left( \begin{array}{c} a_1 \\ a_2 \end{array} \right).
\end{equation}

Furthermore, the operator $\A$ and $\G$ can be symmetrized by setting the complex amplitudes of the eigen-fields $\textbf{E}_1$ and $\textbf{E}_2$ ($\textbf{H}_1$ and $\textbf{H}_2$) to be real (imaginary). Using (\ref{useful}), (\ref{Simp1}) simplifies to
\begin{equation}
\label{DSRR_simp}
\left(\begin{array}{cc}
\omega_1^2 &\-j\omega_1 \frac{\N_{21}}{\A_{11}}\\
-j\omega_2 \frac{N_{12}}{\A_{22}}&\omega_2^2

\end{array}\right)
\left(\begin{array}{c}
a_1\\a2
\end{array}\right)
=\omega^2\left(\begin{array}{c} a_1\\a_2\end{array}\right).
\end{equation}

Solving (\ref{DSRR_simp}), the coupled frequencies are found to be

 \begin{equation}\label{freq_DSRR1}
            \omega_{++}^{2}=\frac{\omega_{1}^{2}+\omega_{2}^{2}}{2}- \sqrt{\left(\frac{\omega_{1}^{2}-\omega_{2}^{2}}{2}\right)^{2}-\frac{\omega_{1}\omega_{2}(\N_{12}\N_{21})}{\A_{11}\A_{22}}}
 \end{equation}
 and
  \begin{equation}\label{freq_DSRR2}
            \omega_{+-}^{2}=\frac{\omega_{1}^{2}+\omega_{2}^{2}}{2}+ \sqrt{\left(\frac{\omega_{1}^{2}-\omega_{2}^{2}}{2}\right)^{2}-\frac{\omega_{1}\omega_{2}(\N_{12}\N_{21})}{\A_{11}\A_{22}}},
        \end{equation}
  where $\omega_{++}$ and $\omega_{+-}$ are the angular frequencies of the symmetric (bonding) and anti-symmetric (anti-bonding) modes respectively. The \emph{strength} of coupling can be quantified by defining the coupling coefficient $\kappa$ as

        \begin{equation}\label{Kappa}
            \kappa(\omega_{1}, \omega_{2})=\sqrt{-\frac{{\N_{12}\N_{21}}}{\omega_{1}\omega_{2}\A_{11}\A_{22}}}.
        \end{equation}

 Due to the interaction (off-diagonal terms) between the uncoupled modes, the coupled frequencies $\omega_{++}$ and $\omega_{+-}$ are different from $\omega_1$ and $\omega_2$. The interaction is due to the $\langle \textbf{E}_i|\textbf{J}_k\rangle$ term. This power interaction can be also explained in terms of the electric and magnetic field overlaps as given by (\ref{useful}) and (\ref{Simp1}). An important special case is when $\omega_1=\omega_2=\omega_0$. This, for example, represents the meta-dimer studied in \cite{Powell10, Liu09} using magnetic and electric dipole interactions. For meta-dimers, (\ref{Simp1}) simplifies to
 \begin{equation}
\label{Degen}
\left(\begin{array}{cc}
\omega_0^2 &\omega_0^2\frac{\A_{12}-\G_{12}}{\A_{11}}\\
\omega_0^2\frac{\A_{21}-\G_{21}}{\A_{22}} & \omega_0^2
\end{array}\right)
\left( \begin{array}{c} a_1 \\ a_2 \end{array} \right)=\omega^2\left( \begin{array}{c} a_1 \\ a_2 \end{array} \right)
\end{equation}

Eq.~(\ref{Degen}) says that $\kappa$ can be regarded as the difference between a magnetic ($\kappa_m$) and an electric ($\kappa_e$) components. This result was previously derived using the Lagrangian equation of motion \cite{Powell10} and the perturbation method \cite{Awai206}. It is also consistent with lumped circuit models, where coupling is modelled by a mutual inductance $\left(L_m={\G_{12}}/{\sqrt{\A_{11}\A_{22}}}\right)$ and a mutual capacitance $\left(C_m={\A_{12}}/{\sqrt{\A_{11}\A_{22}}}\right)$ \cite{Hong11}. When $\kappa_m=\kappa_e$, the modes decouple. This reinforces the findings of \cite{Powell10} that there is no avoided crossing whenever $\kappa_m=\kappa_e$ as was before attributed to the higher order electric multipolar interactions \cite{Liu09}. The decoupling of modes even though $\kappa_m$ and $\kappa_e$ are not negligibly small may seem counter intuitive. Visualizing the coupling as a hybridization of two atomic structures at which the electric-electric dipole and the magnetic-magnetic dipole interactions counteract, alleviate the confusion. The eigenvalue problem (\ref{DSRR_simp}) gives an alternative physical explanation by taking advantage of (\ref{Kappa}). The condition $\kappa_m=\kappa_e$ is equivalent to $\N_{ik}=0$ or  equivalently $\int_{V} \textbf{E}_i\cdot\textbf{J}_k^*dv=0$. This means that there is no interaction between the uncoupled modes, whenever the relative position and orientation of the meta-dimer atoms were meticulously tuned such that $\textbf{E}_i$ is orthogonal to $\textbf{J}_k $. In another words, there is no energy transfer, or a pathway, between the two uncoupled modes and hence no split in frequency.

\begin{figure}[h!]

  \centering
   \includegraphics[width=8cm]{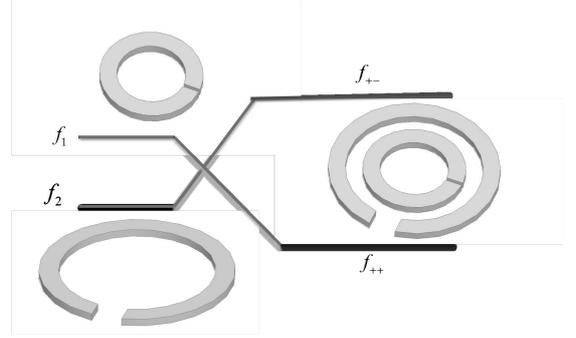}
   \caption{Configuration \emph{A}: A circular double split ring resonator. The frequency of the small ring (resonator 1): $f_1=15$ GHz. That of the large ring (resonator 2): $f_2= 10.3$ GHz. The right hand side shows the coupled frequencies; Symmetric (Bonding): $f_{++}$ and Anti-symmetric (Anti-bonding): $f_{+-}$.}

\label{CircHybrid}
\end{figure}

In the following, two configurations, \emph{A} and \emph{B}, are treated separately. Configuration \emph{A} consists of two coaxial circular SRR which have resonant frequencies of 10.3 GHz and 15 GHz (Fig.\ref{CircHybrid}). The net coupling strength (reflected in the magnitude of the frequency split) is calculated as a function of the angle $\theta$ between the two gaps.

\begin{figure}[h!]

  \centering
   \includegraphics[width=8cm]{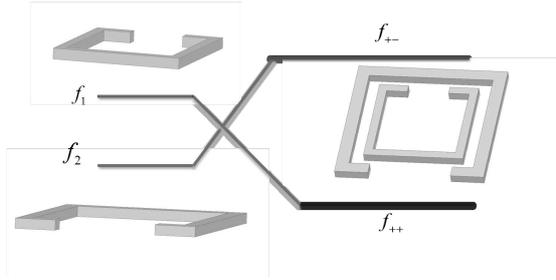}
   \caption{Configuration \emph{B}: A rectangular double split ring resonator. The frequency of the outer ring (resonator 2) is $f_2=1.35\ \textnormal{THz}$. The capacitive gap of the inner ring $W_g=2-16 \ \mu \textnormal{m}$. The right hand side shows the coupled frequencies; Symmetric (Bonding): $f_{++}$ and Anti-symmetric (Anti-bonding): $f_{+-}$.}

\label{RectHybrid}
\end{figure}

Fig.~(\ref{RectHybrid}) shows Configuration \emph{B}. It consists of two co-axial square split rings. The outer ring has a fixed capacitive gap and hence a fixed resonant frequency of $f_2=1.35\ \textnormal{THz}$. The inner ring's gap is allowed to change from 2 $\mu \textnormal{m}$ to 16 $\mu \textnormal{m}$ which is translated to a frequency range of $f_1= 1.64-2.07\ \textnormal{THz}$.

 The two configurations \emph{A} and \emph{B} are quantitatively studied by solving the eigenvalue problem (\ref{eigenDSRR}). As a first step and due to the lack of analytical expressions, the fields and frequencies of the single uncoupled SRRs are computed using HFSS\textsuperscript{\textregistered} eigenmode solver (Ansys Corporation, Pittsburgh, PA, USA). The fields are exported to a MATLAB\textsuperscript{\textregistered} code where the matrices $\G$ and $\A$ are calculated and hence (\ref{eigenDSRR}) is solved to determine the coupled frequencies. Finally, the frequency values are compared to the ones obtained by another HFSS eigenmode simulation of the complete DSRR systems. For the HFSS calculations, the conductivity of the SRR was assumed to be infinite. The solution domain was enclosed in an airbox which is 7 times larger than the SRR width. The structure is considered to be embedded in open space. Therefore, the airbox was subjected to a Perfectly Matched Layer (PML) boundary condition \cite{Gedney96}.

\subsubsection{Analysis of Configuration \emph{A}}

\begin{figure}[h!]

  \centering
   \includegraphics[width=8cm]{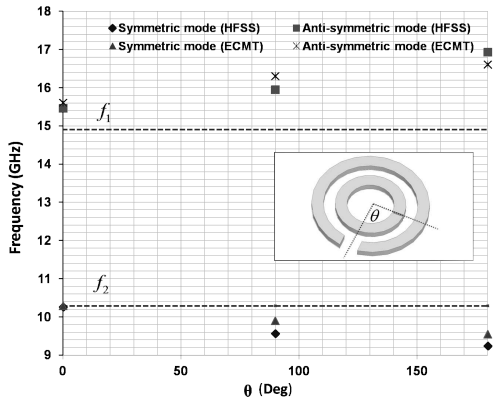}
   \caption{Circular DSRR: The frequencies of the symmetric ($f_{++}$) and anti-symmetric ($f_{+-}$) modes for different $\theta$ values.}

\label{freqCircDSRR}
\end{figure}

\begin{figure}[h!]

  \centering
   \includegraphics[width=8cm]{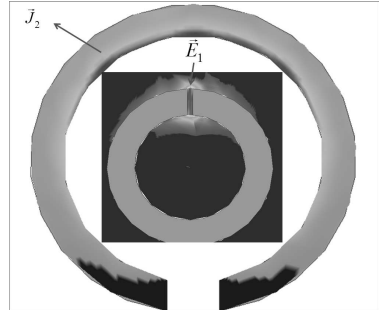}
   \caption{Circular DSRR: The plot of complex magnitude of the Electric field of the inner ring ($\textbf{E}_1$) superimposed on the plot of the magnitude of the surface current density ($\textbf{J}_2$).}

\label{CSRR_OV}
\end{figure}

Fig.~(\ref{freqCircDSRR}) shows the calculated frequency of configuration \emph{A} using both (\ref{eigenDSRR}) and HFSS eigenmode solver. It is clear from the figure that as the angle $\theta$ increases, the coupled frequencies deviate more from the uncoupled ones ($f_1$ and $f_2$). This can be explained by referring to coupling coefficient expression (\ref{Kappa}). Both uncoupled angular frequencies $\omega_1$ and $\omega_2$ are constant. The interactions $\N_{12}$ and $\N_{21}$ are the only terms that change. Thus, $\kappa$ is always proportional to the reactive powers $\int_{\C_k} \textbf{E}_i\cdot \textbf{J}_k^* dv,$ where $\C_k$ is the surface of the $k^{th}$ ring. The electric field of the uncoupled modes is concentrated in the gap of the SRR. At the same time the conduction current $\textbf{J}_k$ attains its maximum at the farthest side. Therefore when $\theta$ increases, $\kappa$ increases. Fig. (\ref{CSRR_OV}) clarifies this by superimposing the calculated electric field distribution of the inner SRR ($\textbf{E}_1$) on the same plot of the calculated magnitude of the current density ($\textbf{J}_2$) of the outer SRR when $\theta=180^{\circ}$. Clearly as $\theta$ decreases, $\N_{ik}$ decreases and so $\kappa$.

\subsubsection{Analysis of Configuration \emph{B}}

\begin{figure}[h!]

  \centering
   \includegraphics[width=8cm]{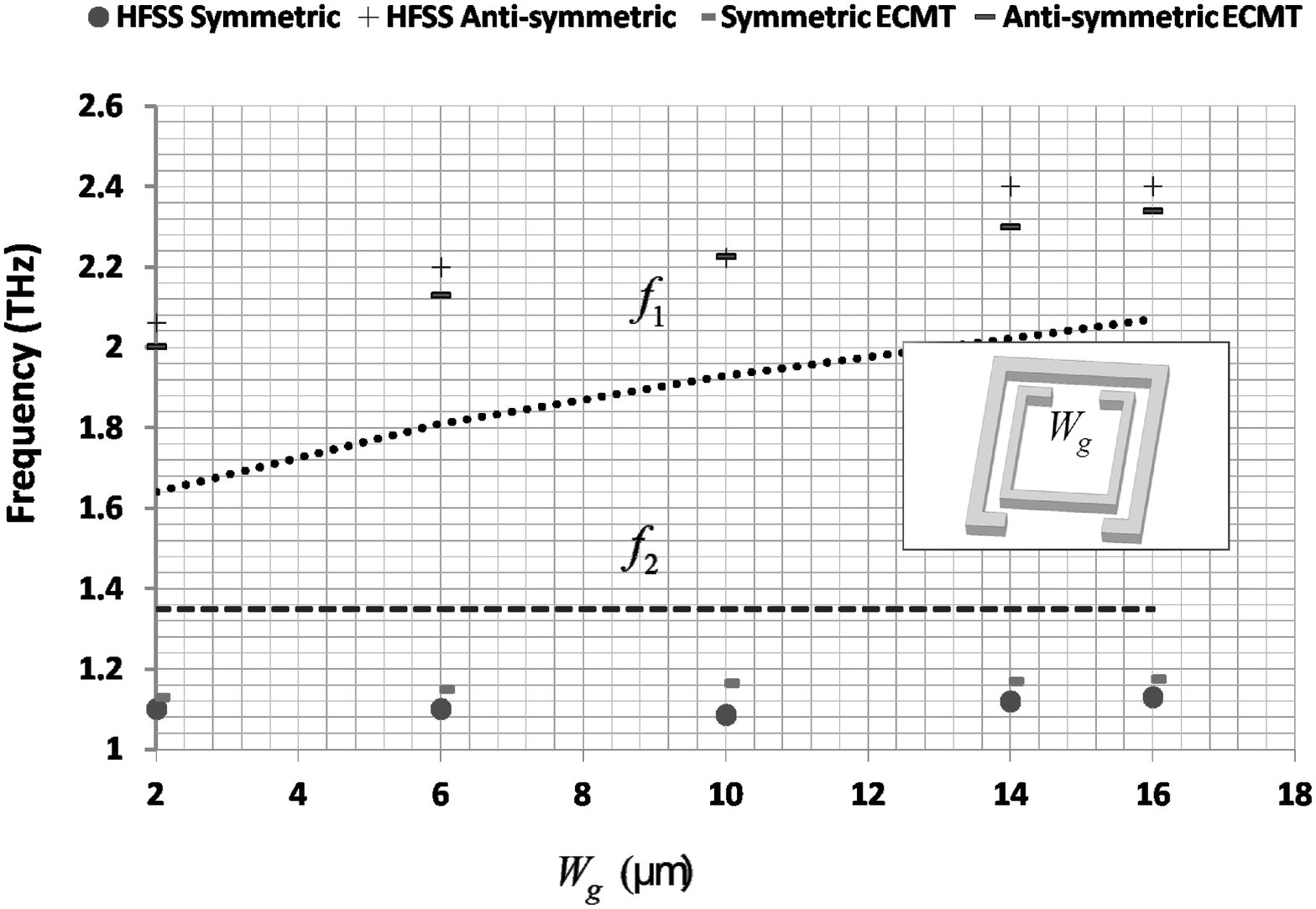}
   \caption{The frequencies of the symmetric ($f_{++}$) and anti-symmetric ($f_{+-}$) modes of the rectangular double split-ring resonator.}

\label{freqRectDSRR}
\end{figure}

\begin{figure}[h!]

  \centering
   \includegraphics[width=8cm]{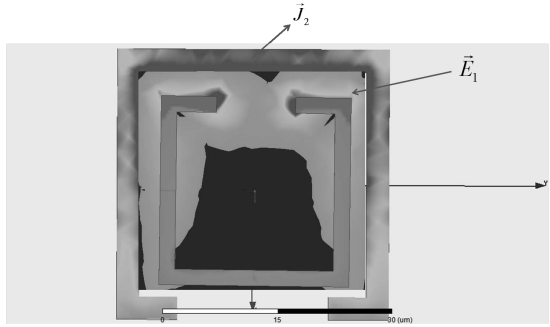}
   \caption{Rectangular DSRR: The plot of complex magnitude of the Electric field of the inner ring ($\textbf{E}_1$) superimposed on the plot of the magnitude of the surface current density ($\textbf{J}_2$). When the capacitive gap $W_g$ of the inner ring is extended, $\textbf{E}_1$ spreads across a larger area. }

\label{SRR_OV1}
\end{figure}

Configuration \emph{B} is more interesting. Not only are the resonant frequencies in the far infrared, but also the uncoupled frequency $\omega_1$ does change. Accordingly and from (\ref{Kappa}), $\kappa$ is now a function of both the interaction terms $\N_{ik}$ and the frequency $\omega_1$. The calculated frequencies are presented in Fig. (\ref{freqRectDSRR}) where again the values computed using (\ref{eigenDSRR}) are compared to those obtained by HFSS eigenmode solver. The results confirm the applicability of (\ref{eigenDSRR}). It is also observed that as $\omega_1$ increases, the shift in frequency of the anti-symmetric mode $\mid\omega_{+-}-\omega_1\mid$ decreases. From (\ref{freq_DSRR2}), the frequency shift is a function of the product $\omega_1|\N_{12}|$, which, as estimated in the \emph{Appendix}, decreases whenever $W_g$ increases. To clarify why $|\N_{12}|$ decreases when $W_g$ increases, one refers to Fig. (\ref{SRR_OV1}) which illustrates how the hybridization can be visualized in terms of the interaction between $\textbf{E}_1$ and $\textbf{J}_2$ of the uncoupled modes. From the figure the current density values $\textbf{J}_2$ are maximum near the inner ring's gap. Therefore when $W_g$ increases ($\omega_1$ increases), $\textbf{E}_1$ is distributed over a larger width and thus reduces $\N_{12}$. It is worth noticing that the frequency shift $\mid \frac{\omega_{+-}-\omega_1}{\omega_1}\mid$ is significantly large ($\approx 15-25\%$).

Unlike $\omega_{+-}$, $\omega_{++}$ does not significantly change with $W_g$. This cannot be explained by simply referring to (\ref{freq_DSRR1}) which was derived based on the assumption that higher order terms are negligibly small. In fact, (\ref{freq_DSRR1}) and (\ref{freq_DSRR2}) predict that
\begin{equation}
\label{omega-omega}
\omega_{+-}^2-\omega_{1}^2=\omega_2^2-\omega_{++}^2,
\end{equation}
which does not comply with the curves depicted in Fig.~(\ref{freqRectDSRR}). To better understand why $\omega_{++}$ behaves as shown in Fig.~(\ref{freqRectDSRR}), the higher order terms in the on-diagonal elements are retained. The on-diagonal terms are modified by subtracting  $\chi_{ii}=\omega_1\omega_2 \frac{\G_{ik}\A_{ki}}{\A_{11} \A_{22}}$ from $\omega_i^2$. For the DSRR shown in Fig.~(\ref{RectHybrid}), $\A_{ki}<0$ and hence $\chi_{ii}<0$ . This is because the angle between $\textbf{E}_1$ and $\textbf{E}_2$ is $180^\circ$. $\chi_{ii}$, the coupled induced frequency shift coefficient, was theoretically described for coupled optical cavities \cite{Popovic06}. The expressions (\ref{freq_DSRR1}) and (\ref{freq_DSRR2}) for the coupled frequencies are then modified by replacing each $\omega_i^2$ with $\omega_i^2-\chi_{ii}$. Accordingly, (\ref{omega-omega}) becomes
\[
\omega_2^2-\omega_{++}^2=\omega_{+-}^2-\omega_1^2+\chi_{11}+\chi_{22},
\]
(i.e.,the shift between $\omega_{++}$ and $\omega_2 $ is smaller than that between $\omega_{+-}$ and $\omega_1$.) Thus, the effect of $\chi_{ii}$ is to \emph{pull} $\omega_{++}$ up toward $\omega_{2}$ and counteracts the influence of the off-diagonal cross coupling term. With a similar argument to the one presented in the \emph{Appendix}, it can be shown that $|\chi_{ii}|$ decreases as $W_g$ increases, which keeps $\omega_{++}$ curve approximately flat as Fig.~(\ref{freqRectDSRR}) shows.

\begin{figure}[h!]

  \centering
   \includegraphics[width=8cm]{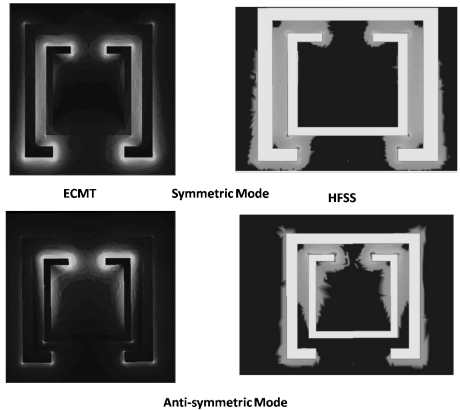}
   \caption{The electric fields of the coupled modes of the rectangular double split-ring resonator when $W_g=10\ \textnormal{m}\mu$ ($f_1\approx=1.93\ \textnormal{THz}$). (a) The electric field of the symmetric mode. (b) The electric field of the anti-symmetric mode.}

\label{SRR_Fields}
\end{figure}

To determine the fields using the coupled mode formalism, the eigenvectors for the coupled modes are computed and the expansion (\ref{EfieldExp}) is used. Fig. (\ref{SRR_Fields}) shows the electric field of configuration \emph{B} when $W_g=10\ \mu \textnormal{m}, f_1\approx 1.93\ \textnormal{THz}$. Because the total electric field does not satisfy the boundary conditions at the rings' surface, the field calculated is not exact. Nevertheless, the eigenvalue problem (\ref{eigenDSRR}) still gives very reasonable results as it shows the contribution of each of the uncoupled modes to the total DSRR fields.

\section{Conclusion}
A general coupled mode equation in the form of an eigenvalue problem is derived. The eigen-frequencies are determined after finding the eigenvalues. The eigenvectors are used to find the electromagnetic fields. If resonators are compared to atoms, the eigenvalue problem can be considered as the electromagnetic analog of molecular orbital theory. This conceptual view agrees with the way meta-materials unit cells are treated. It is shown that the eigenvalue equation obeys the energy conservation principle. As an immediate application, the behavior of meta-dimers and DSRR was explained using the interaction between $\textbf{E}_i$ and $\textbf{J}_k$. Thus, the eigenvalue problem provides an intuitive view to how resonators interact. The $\left(\textbf{E}_i,~\textbf{J}_k\right)$ interaction picture is equivalent to other well known methods of analysis such as the dipole interactions and lumped circuit models. Two configurations were formulated and numerically solved and the results were compared to finite element simulations. To illustrate the versatility of the coupled mode formalism, the numerical findings were explained using the $\left(\textbf{E}_i,~\textbf{J}_k\right)$ interaction picture. It was shown that the coupled induced frequency shifts terms $\chi_{ii}$ is very essential to correctly explain and quantify the DSRR behavior.

\section*{Appendix}
Consider a simple LC circuit, with a capacitive gap $W_g$, resonating at angular frequency $\omega_1$. In terms of the voltage on the capacitor $C$, the average power is
\[
P=\frac{1}{2}\sqrt{\frac{C}{L}}V^2,
\]
where the relation $\omega_1=\frac{1}{\sqrt{LC}}$ was used. The capacitance $C\propto 1/W_g$, $V\propto |\textbf{E}|W_g$. Thus, for a fixed power $P$,
\[
\omega_1|\textbf{E}|\propto \frac{1}{\sqrt[4]{W_g}}.
\]
$\N_{12}$ is the integral of $\textbf{E}_1$ and $\textbf{J}_2$. Therefore $\omega_1|\N_{12}|\sim \frac{1}{\sqrt[4]{W_g}}$


%\begin{IEEEbiographynophoto}{Richard Tervo}

%\newpage



%\vfill


\end{document}